\documentclass[oldversion]{aa}
\newcommand{\vx}{\mbox{\bf {x}}}

\newcommand{\vk}{\mbox{\bf {k}}}

\newcommand{\vv}{\mbox{\bf {v}}}

\newcommand{\vnh}{\hat{\mbox{\bf {n}}}}

\usepackage{multirow}
\def\plotone#1{\centering \leavevmode       
\epsfxsize=\columnwidth \epsfbox{#1}}         

\usepackage{graphicx}
\usepackage{txfonts}
\usepackage{natbib}
\usepackage{epsfig}

\begin{document}
   \title{Galaxy Clusters as mirrors of the distant Universe}
   \subtitle{Implications for the kSZ and ISW effects}

   %\subtitle{I. Overviewing the $\kappa$-mechanism}

   \author{Carlos Hern\'andez--Monteagudo
          \inst{1}
          \and
          Rashid A. Sunyaev\inst{1,2}\fnmsep
                    }

   \institute{Max Planck Institut f\"ur Astrophysik, 
              Karl Schwarzschild Str.1, D-85741,
              Garching bei M\"unchen, Germany\\
              \email{chm@mpa-garching.mpg.de}
          \and
          	Space Research Institute, Russian Academy of Sciences, Profsoyuznaya 84/32, 117997 Moscow, Russia
             \\
             \email{sunyaev@mpa-garching.mpg.de}
                          }

   \date{Received ; accepted }

% \abstract{}{}{}{}{} 
% 5 {} token are mandatory
 
  \abstract
  % context heading (optional)
  % {} leave it empty if necessary  
%   {investigate.}
  % aims heading (mandatory)
%   {It is shown that }
  % methods heading (mandatory)
%   {The stability equations }
  % results heading (mandatory)
 %  {Vibrational instability }
  % conclusions heading (optional), leave it empty if necessary 
 %  {}
 %__ traditional abstract
 {
 It is well known that Thomson scattering of Cosmic Microwave Background (CMB) photons in galaxy clusters introduces new anisotropies in the CMB radiation field, but however little attention is  payed to the fraction of CMB photons that are scattered {\em off} the line of sight,  causing a slight blurring of the CMB anisotropies present {\em at the moment} of scattering. In this work we study this {\it blurring} effect, and find that it has a non-negligible impact on estimations of the kinetic Sunyaev-Zel'dovich (kSZ) effect: it induces a 10\% correction in 20-40\% of the clusters/groups, and a  100\% correction in $\sim 5$\% of the clusters in an ideal (noiseless) experiment. We explore the possibility of using this blurring term to probe the CMB anisotropy field at different epochs in our Universe. In particular, we study the required precision in the removal of the kSZ that enables detecting the blurring term $-\tau_T \delta T / T_0$ in galaxy cluster populations placed at different redshift shells.  By mapping this term in those shells, we would provide a tomographic probe for the growth of the Integrated Sachs Wolfe effect (ISW) during the late evolutionary stages of the Universe.  We find that the required precision of the cluster peculiar velocity removal is of the order of 100 -- 200 km s$^{-1}$ in the redshift range 0.2 -- 0.8, after assuming that all clusters more massive than 10$^{14}$ h$^{-1}$ M$_{\odot}$ are observable. These errors are comparable to the total expected  linear line of sight velocity dispersion for clusters in WMAPV cosmogony, and correspond to a residual level of roughly 900 -- 1800 $\tau_T \mu$K per cluster, including all types of contaminants and systematics. Were this precision requirement achieved, then independent constraints on the intrinsic cosmological dipole would be simultaneously provided.
 }

   \keywords{(Cosmology) : cosmic microwave background, Large Scale Structure of the Universe
                             }
\titlerunning{Galaxy clusters as mirrors of distant Universe}

   \maketitle
%
%________________________________________________________________

\section{Introduction}

The study of the scattering of Cosmic Microwave Background (CMB) photons in
moving clouds of thermal electrons (like galaxy clusters) has been a subject of active research since the
first works of \citet{sz70,tSZ,kSZ,sz1}. Indeed, this mechanism is one of the key two
sources for intensity and polarization anisotropies in the CMB (the other one
being associated to the presence of gravitational fields, the so-called Sachs-Wolfe effect, \citep{SW}). 
The Thomson
scattering changes the angular distribution of the CMB photons, partially
erasing the original anisotropy pattern (due to the {\em off} scattering of
photons propagating initially along the line of sight) and introducing new
anisotropies if the electrons move with respect to the CMB rest frame. Due to
its anisotropic nature, Thomson scattering also introduces linear polarization
if the CMB shows an intensity quadrupole at the scattering place.  The use of
the polarization induced by Thomson scattering in galaxy clusters has been
proposed as a probe for remote CMB quadrupoles, \citep{kSZ,marcavi,sazonovrashidpcl,challinorp},
with implications for the Integrated Sachs-Wolfe effect
\citep[ISW,][]{cooraypisw} and for the characterization of the large scale
density distribution in the observable universe, \citep[e.g.,][]{setop,abramop}\\

However, this work will be devoted to the study of a  {\it
  blurring} term, i.e., the term responsible for the smearing of the original
intensity anisotropies of the radiation field. For an axisymmetric radiation
field described by $T(\mu )/T_0 = \sum_l a_l P_l(\mu) $ (with $\mu \equiv \cos \theta$ measuring deviations from the symmetry axis), the blurring term
will affect {\em all} multipoles, whereas the anisotropic scattering will
slightly modify the quadrupole ($a_2$) 
\citep[e.g.][]{kSZ,sz1,SZreview}:
\begin{equation}
\frac{\Delta T}{T_0} = \tau_T\;\left(-\sum_{l=1}^{\infty}a_l P_l(\mu) +
\frac{1}{10}a_2P_2(\mu)\right),
\label{eq:rt0}
\end{equation} 
where $a_{l=1}$ refers to the intrinsic {\em cosmological} dipole at the scattering place.
However, if the scatterers show a peculiar motion with respect to the radiation field, an additional
Doppler term must be added to the previous expression:
\begin{equation}
\frac{\Delta T}{T_0} = \tau_T\;\left(-\sum_{l=1}^{\infty}a_l P_l(\mu) +
\frac{1}{10}a_2P_2(\mu) - \frac{\vv\cdot \vnh}{c} \right).
\label{eq:rt0}
\end{equation} 
with $\vnh$ the unit
vector along the line of sight and $\vv$ the scatterer's proper peculiar
velocity. Since $\vv\cdot\vnh/c$ is typically larger than $\Delta T/T_0$
($\vv\cdot\vnh /c \simeq 7\times 10^{-4}$ in the linear approximation for WMAPV
cosmology, $\Delta T/T_0 \sim 10^{-5}$ as measured by COBE and WMAP), the
blurring term should {\it a priori} be a typically few percent correction to the peculiar motion
effect, \citep[hereafter the kinetic effect, kSZ,][]{kSZ}. However, this ratio does depend on the cluster's peculiar motion and angular size, as we shall show later. At any rate, we show that this blurring term has potentially relevant
cosmological significance: apart from the impact it has on accurate measurements of the kSZ effect, it can also provide an alternative measurement of the gas context in clusters. Likewise, if the Thomson scattering optical depth associated to the cluster is known, by using CMB observations (like those of WMAP\footnote{WMAP's URL site:\\ {\tt http:lambda.gsfs.nasa.gov/product/map/current/}} or Planck\footnote{Planck's URL site: \\{\tt http://www.esa.int/esaMI/Planck/index.html}}) this effect can be accurately predicted as an attenuation of the CMB anisotropy in the direction of the cluster. 

If detected on a set of cluster samples placed at different redshifts, this blurring term can be used as a probe {\it in situ} of the CMB anisotropy field at those epochs. By looking at the variations of this term at different redshift shells, one should 
be able to track the growth of new anisotropies arising at later times. In particular, it should enable performing tomography of the ISW effect, generated by the decay of the linear gravitational potentials at late epochs. 

This acquires particular relevance in the context of future surveys like the X-ray mission SPECTRUM-X/eROSITA\footnote{Spectrum-X/eROSITA's URL site: \\{\tt
    http://www.mpe.mpg.de/projects.html\#erosita}}, whose prospects is to locate $\sim 150,000$ galaxy clusters on the sky, many of which will sit on top of high amplitude CMB intensity excursions where this effect can be measured more easily. Furthermore, we also show that this phenomenon should also provide stringent
limits on the amplitude of the intrinsic cosmological dipole. \\

This paper is organized as follows: in Section 2 we describe the blurring term
within the context of Thomson scattering. In Section 3 we study its implications in the measurement of kSZ  and remote quadrupoles at the position of galaxy clusters. In Section 4 we introduce the possibility of using the blurring term for tracking the growth of the ISW: we analyze the requirements in the cluster
sample and in the peculiar velocity recovery. We observe the possibility of setting constraints on the cosmological dipole by using this effect in Section 5. Finally, in Section 6 we discuss our results and conclude.

\section{The scattering of CMB photons in electron clouds}
 
%Describe the radiative transfer equation. All terms. Stress the absorption term in the intensity
%and polarization.
%Show how the C_l's seen by the cluster depend on the redshift. Also the polarization! Include a plot for the extra volume probed by the quadrupole at the cluster population.

Thomson scattering modifies the angular pattern of the CMB intensity and polarization anisotropies. The source for new intensity anisotropies is associated to the peculiar velocity of the gas cloud with respect to the CMB frame \citep{kSZ}, whereas in the case of polarization anisotropies is exclusively associated to the CMB quadrupole at the scattering place. If the electron gas is at high temperature, then Compton scattering transfers energy from the electron plasma to the CMB photon field, distorting the CMB black body spectrum and introducing {\em frequency dependent} temperature fluctuations \citep[ hereafter denoted as tSZ effect]{tSZ}. The tSZ effect (and its relativistic corrections) has a definite spectral dependence \citep{tSZ,SZreview,reltSZ}, so hereafter we shall assume that it can be accurately subtracted. With this in mind, the change in the CMB intensity and polarization pattern when it crosses an electron cloud can be written as \citep{Jmethod}:

\begin{eqnarray}
\Delta \left[ \frac{\delta T}{T_0}\right] (\vnh ) & = & -\tau_T \frac{\vnh\cdot \vv}{c}-\tau_T \frac{\delta T}{T_0}(\vnh ) \biggr|_{cl} +\\
\label{eq:rt1}
\nonumber
& &  \tau_T \; \int d\vnh' \biggl({\cal M}_{T,T} (\vnh,\vnh') \frac{\delta T}{T_0}(\vnh' )\biggr|_{cl} +\\
\nonumber
& &
\phantom{xxxx}
{\cal M}_{T,+} (\vnh,\vnh')(Q+iU)(\vnh' )\biggr|_{cl} +\\
\nonumber
& & 
\phantom{xxxx}
{\cal M}_{T,-} (\vnh,\vnh')(Q-iU)(\vnh' )\biggr|_{cl} 
\biggr)
\nonumber
%\label{eq:rt1b}
\nonumber
\end{eqnarray}
\begin{eqnarray}
\Delta \left[ Q\pm iU \right] (\vnh ) & = & -\tau_T\;Q\pm iU (\vnh)\biggr|_{cl} \;+\\
& &  
\nonumber
\tau_T \; \int d\vnh' \biggl({\cal M}_{\pm,T} (\vnh,\vnh') \frac{\delta T}{T_0}(\vnh' )\biggr|_{cl} +\\
& &
\nonumber
\phantom{xxxx}
{\cal M}_{\pm,+} (\vnh,\vnh')(Q+iU)(\vnh' )\biggr|_{cl} +\\
& & 
\nonumber
\phantom{xxxx}
{\cal M}_{\pm,-} (\vnh,\vnh')(Q-iU)(\vnh' )\biggr|_{cl} 
\biggr)
\end{eqnarray}
where the $cl$ subscript refers to {\it cluster position} and $\tau_T$ denotes its optical depth. The kernels ${\cal M}_{i,j}$ with $i,j=T,+,-$ express the generation of new polarization/quadrupole at the cluster position, \citep{sazonovrashidpcl}. Note that these expressions are up to second order in the quantities associated to the cluster \citep[$\vv\cdot\vnh$, $\tau_T$,][]{SZreview}. When considering 
higher order terms, new source terms for polarization and intensity in the direction of galaxy clusters arise, observing combinations of terms of the type $\tau_T v_r^2$, $\tau_T v_t^2$ for a single scattering and terms of the type $\tau_T^2 \;v_r$, $\tau_T^2 v_t$ for double scattering, \citep[$v_r$ and $v_t$ denote the radial and tangential components of the scatterer's peculiar velocities with respect to the line of sight, ][]{SZreview,ink}). We remind that, due to their peculiar frequency dependence, we are ignoring all terms associated to corrections of the tSZ and consider only those which are spectrally indistinguishable from the primordial CMB anisotropies.

Our interest in this paper will be focused in the intensity blurring term  ($-\tau_T \delta T(\vnh)/T_0$), which accounts for the fraction of photons that, initially propagating along the line of sight, were scattered {\em off} it and never reach the observer. This term hence describes the {\it erasing} of the CMB anisotropies {\em at the scattering position along the line of sight towards the electron cloud}, since the Thomson scattering tends to {\em isotropize} the CMB angular fluctuations in that direction. The source of polarization, instead, is the {\em local} CMB intensity quadrupole, which is sensitive to the CMB at {\em all} directions in that position. In what follows, let us regard the galaxy cluster and group population as clouds of free electrons.

%{\bf should we say something more about the polarization??}

\section{Impact on kSZ estimations}

\begin{figure}
\centering
%\plotancho{./lM.eps}
\includegraphics[width=9.cm,height=14.cm]{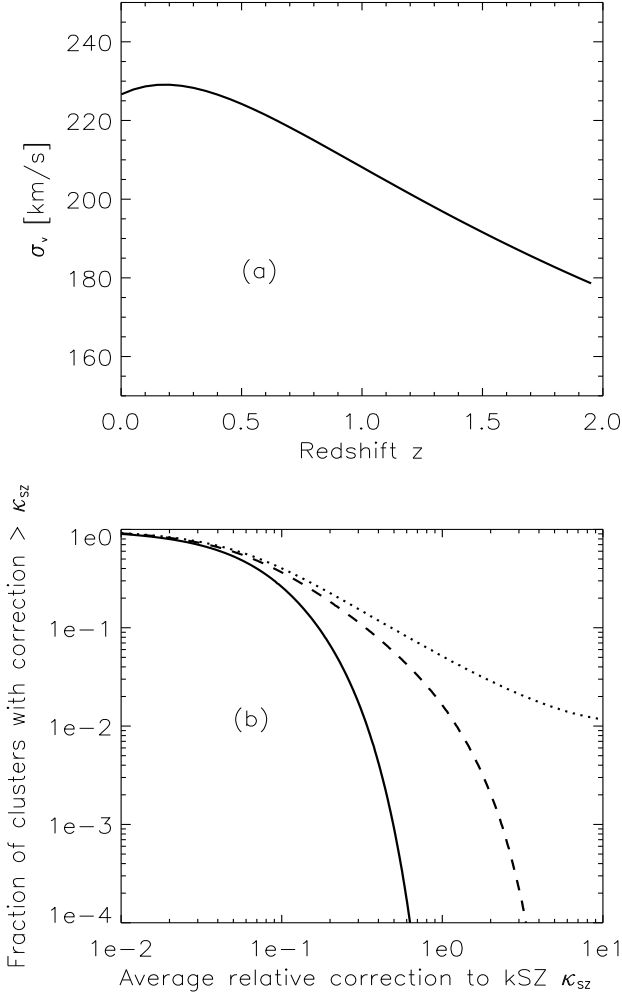}
\caption[fig:pecv1]{{\it (a)} Amplitude of the linear rms radial peculiar velocity in a WMAPV cosmology for a $2\times 10^{14}$ h$^{-1}$ M$_{\odot}$  cluster of galaxies. {\it (b)} Fraction of clusters on the sky above a given fractional level of correction on the kSZ induced by the blurring term. Dotted, dashed and solid lines correspond to a level of residual (instrumental noise related) radial velocity of $60$, $10$ and $0$ km s$^{-1}$.
}
\label{fig:pecv1}
\end{figure}

\begin{figure}
\centering
\includegraphics[width=9.cm,height=9.cm]{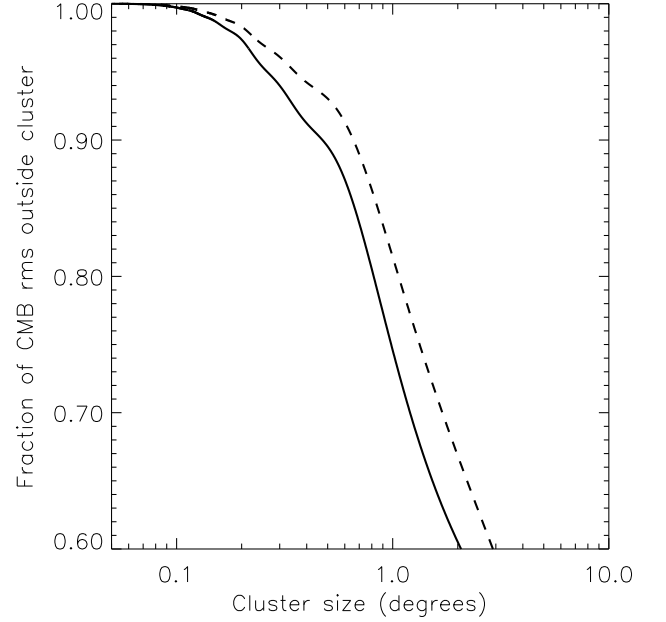}
\caption[fig:sgcmb1]{Fraction of the total CMB temperature rms that is generated outside the cluster size. The solid line considers full sky CMB observations ($l_{min}=2$), the dashed line a patch of roughly 10\degr on a side ($l_{min} = 20$).
}
\label{fig:sgcmb1}
\end{figure}

As mentioned above, the kSZ effect in clusters has the same spectral dependence of the intrinsic CMB anisotropies, and therefore extracting it requires the use of spatial frequency information.
Its amplitude is directly proportional to the projection of the cluster's peculiar velocity along the line of sight. In linear theory, and after assuming that the initial rotational component of the peculiar velocity field is negligible (since it scales with the inverse of the cosmological expansion scale factor), it is possible to relate the Fourier modes of the peculiar velocity with those of the matter density field ($\delta_{\vk}$):
\begin{equation}
\vv_{\vk} = -i H(z) \frac{d{\cal D}_{\delta}(z)}{dz} \frac{\delta_{\vk}}{k^2}\vk,
\label{eq:v1}
\end{equation}
where $H(z)$ is the Hubble function and ${\cal D}_{\delta}(z)$ is the linear growth factor for the density perturbations.  After Fourier transforming into real space, it is easy to prove that the three spatial components of the peculiar velocity are uncorrelated, i.e., the tensor $\langle v_i(\vx) v_j (\vx)\rangle$  is diagonal ($\propto \delta_{ij}$) for $i=1,2,3$\footnote{This is not the case in Fourier space, where different spatial components of $\vv_{\vk}$ are correlated}. A direct consequence of it is that studying the line of sight projected component of the real space peculiar velocity is totally equivalent to studying any of its three components. From eq.(\ref{eq:v1}) it is easy to infer that, in linear theory, velocities will be Gaussian distributed, and each component will have a dispersion given by
\begin{equation}
\sigma_v^2 (M)= \frac{1}{3} \int \frac{d\vk}{(2\pi)^3} H^2(z) \biggl| \frac{d{\cal D}_{\delta}}{dz}\biggr|^2 \frac{P_m(k)}{k^2} |W(kR[M])|^2,
\label{eq:v2}
\end{equation}
where $W(kR[M])$ is the Fourier window function associated to a top hat filter of size given by the linear scale corresponding to the cluster mass $M$, $R = [3\rho_m/(4\pi)]^{1/3}$, and $\rho_m$ the matter comoving density. The linear matter power spectrum is given by $P_m(k)$.  In Figure (\ref{fig:pecv1}a) we display the linear rms amplitude of the radial component of the peculiar velocity versus redshift in the standard $\Lambda$CDM scenario. In an Einstein de Sitter (EdS) universe,  its redshift scaling is $\propto 1/\sqrt{1+z}$, but in a $\Lambda$CDM universe its decay is even more slowly (it drops around a $\sim 20$ \% in the redshift range $z\in [0,2]$). Galaxy clusters and groups where the kSZ can be measured are {\it biased}, peculiar regions in the Universe. Different studies have attempted to quantify whether the peculiar velocities of these objects are actually biased with respect to linear theory predictions. \citet{bbks} found that their velocity should be better approximated by that of the peaks in a Gaussian distribution. \citet{shethdia} conducted a through study of peculiar velocities in simulations, in which they showed that also halo peculiar velocities depend upon the environment density, but even in the extreme cases the change in peculiar velocity would be smaller than $\sim 70\%$ of the average value. \citet{peel06} found an average bias of $\sim 30\%$ of the halo peculiar velocity when compared to linear theory predictions.

We have adopted an average bias of 30\% in the halo peculiar velocities, and estimated the average correction to the kSZ that the blurring term gives rise to. For that, we have assumed that both the CMB temperature anisotropies and the kSZ fluctuations are Gaussian distributed, and computed the fraction of clusters whose kSZ undergoes a correction above a certain level $\kappa_{kSZ} \equiv |\delta_{blurr} / \delta_{kSZ}| = |\delta_{CMB} / [(\vv\cdot\vnh)/c + v_{res}/c] |$ due to the blurring effect:
\[
{\cal P} (>\kappa_{kSZ} ) = \int\int_{|\delta_{blurr}'/\delta_{kSZ}'|> \kappa_{kSZ}}  d\left(\frac{\vv\cdot\vnh}{c}\right)' d\delta_{CMB}' \; \times
\]
\begin{equation}
\phantom{xxxxxxxxxxxxxxxxxxxxxxxxxxxxxx}
 p(\delta_{CMB}')\;  p\left[\left(\frac{\vv\cdot\vnh}{c}\right)' \right].
\label{eq:blurprob1}
\end{equation}
In both cases, the function $p(X)$ denoted Gaussian probability distribution function on the variable $X$. The term $v_{res}/c$ is included as a residual radial velocity term that avoids divergences: the solid line in Figure (\ref{fig:pecv1}b) corresponds to $v_{res} = 60$ km s$^{-1}$,  the dashed line to $v_{res} = 10$ km s$^{-1}$ and the dotted line to a negligible level of $v_{res}$, $v_{res} \simeq 0$ km s$^{-1}$. These residual radial velocities obey to the unavoidable level of noise present at the cluster's position when observed by experiments. The level of $60$ km s$^{-1}$ corresponds roughly to a noise level of $1\; \mu$K. Figure (\ref{fig:pecv1}) shows that more than 20\% (5\%) of the clusters undergo a correction on the kSZ that is larger than 10\% (100 \%).   Obviously, for smaller radial velocities (and smaller values of $v_{res}$), the correction induced is larger. Approximately, the blurring amplitude can be expressed as $|\delta T|_{blurr} \sim |\delta T_{CMB}/\sigma_{CMB}| \;(\tau_T / 10^{-2}) \;\mu$K. 

Once the background CMB temperature anisotropies and the gas content of the cluster are known, then the blurring effect can be accurately predicted. Its amplitude will be larger for those regions hosting larger excursions of the CMB intensity field, and should dominate over the kSZ for those clusters having little radial relative motion with respect to the CMB. In the full sky, for WMAPV universe, the CMB intensity rms amplitude is $\sigma_{CMB} \simeq 114 \; \mu$K. This means that, for a survey of $\sim 150,000$ clusters (as the one to be conducted by SPECTRUM-X/eROSITA), around 10 clusters will show decrements of amplitude larger than $470\tau_T$ $\mu$K (4-$\sigma$), and $\sim 400$ clusters will be above the level of $350\tau_T$ $\mu$K, (3-$\sigma$).

When averaging over the cluster's area,  it is of relevance to know what is the amplitude of the rms CMB fluctuations that is being blurred by the cluster. Figure (\ref{fig:sgcmb1}) shows the fraction of this total rms that is being blurred by a cluster of a given size, i.e., the fraction of the total rms of a typical patch of the sky whose size is equal to the cluster size. This would be the effective CMB amplitude that is partially erased if, when pursuing the detection of secondary effects at the cluster's position, the CMB is integrated within the cluster's area. The solid line provides the case when all CMB sky available and $l_{min}=2$, the dashed line considers the case when only a limited patch is available, so that $l_{min} = 20$. In the first case, most of the anisotropy is coming from large angular scales, so the cluster size has little impact on the blurred intensity. For small surveys ($l_{min}=20$ corresponds to a patch size of $\theta \sim 10$\degr on a side), the rms CMB fluctuation has smaller amplitude and the cluster size becomes more important. 

We should also remark that the detection of this effect also provides an independent measurement of the gas content of the cluster, since the ratio of the blurred to the original CMB pattern is simply the cluster's Thomson optical depth $\tau_T$. 

\section{Implications for the ISW effect}

\begin{figure}
\centering
\includegraphics[width=9.cm,height=12.cm]{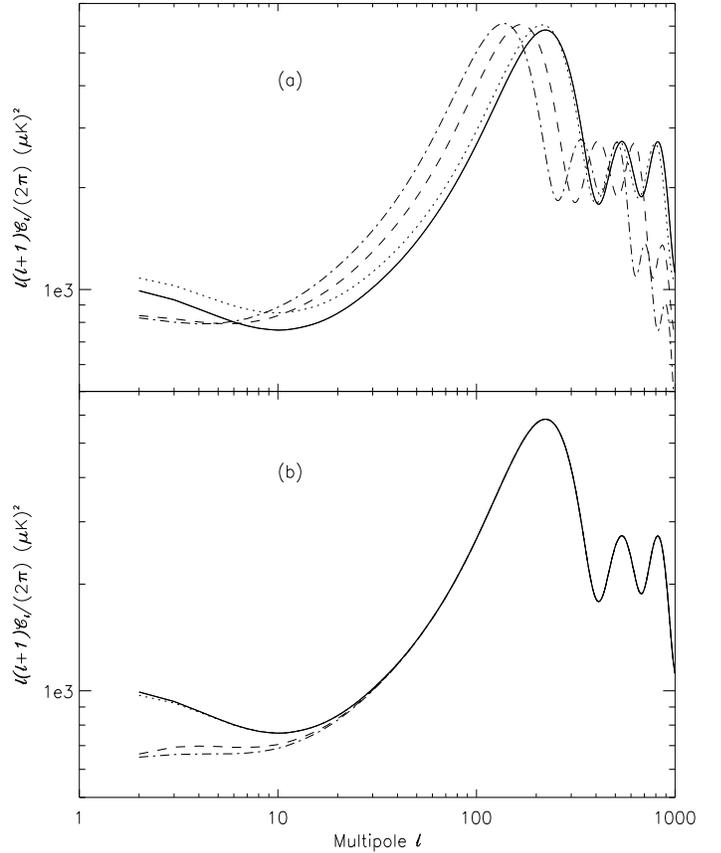}
%\plotancho{./cls_vs_z.eps}
\caption[fig:cls]{{\it (a)} CMB TT angular power spectrum as seen by observers placed at redshifts $z=0, 0.1, 1$ and $2$ (solid, dotted, dashed and dot-dashed lines, respectively). {\it (b)} Same power spectra as in {\it (a)} after being free streamed to the present moment.
}
\label{fig:cls}
\end{figure}

\begin{figure}
\centering
%\plotancho{./lM.eps}
\includegraphics[width=9.cm,height=8.cm]{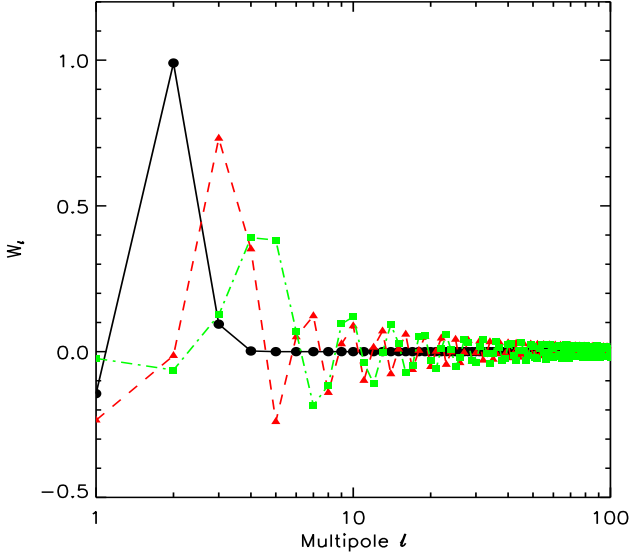}
\caption[fig:wdwquad]{ Proyection of the CMB quadrupole multipole $a_{2,0}$ as seen by observers placed at different redshifts into different $a_{l,0}$ multipoles observed at present. Filled circles connected with a black solid line correspond to an observer placed at $z=0.1$, red triangles by a red dashed line to an observer at $z=1$, and green squares linked by dot-dashed line to $z=2$.
}
\label{fig:wdwquad}
\end{figure}

By measuring the intensity blurring term through the lines of sight towards clusters placed at a given redshift shell, then an estimate of the CMB temperature field {\em at that redshift} would be obtained. In Figure (\ref{fig:cls}a) we show the CMB TT (intensity) power spectrum as measured by observers placed at different redshifts; thick solid line corresponds to $z=0$, dotted line to $z=0.1$, dashed line to $z=1$ and dot-dashed line to $z=2$. Since those observers are closer to the surface of last scattering, the whole acoustic pattern shifts to larger angular scales (smaller multipoles). Furthermore, at redshifts larger than $z\sim 1$ the contribution of the ISW is very small, and this is also visible in the low $l$ range. After the scattering, the angular pattern of the CMB anisotropies would be free streamed towards the observer, shifting the whole picture "back" to its standard position, (see Figure (\ref{fig:cls}b)).
 In Figure (\ref{fig:wdwquad}) we display the free streaming of the CMB quadrupole multipole  $a_{2,0} (\vnh )$ as seen by an observer placed at different redshifts into different multipoles $a_{l,0}$-s. The case of redshift $z=0.1$  is displayed by solid circles joined by a solid black line, $z=1$  by red triangles joined by a dashed line, and $z=2$ by green squares joined by a dot-dashed line. The further the remote observer is, the more power is aliased into high $l$ multipoles. This streaming of CMB angular anisotropies permits comparing the ISW pattern at different cosmological epochs on the angular/multipole scale.
If the blurring term is observed through the line of sights corresponding to a population of galaxy clusters and groups situated at large redshift, then it would provide a picture of the CMB pattern {\em before} the ISW arises. I.e., by observing the blurring term in galaxy clusters placed at different redshift shells it should be possible, a priori, to track the growth of the ISW effect with decreasing redshift. 

\begin{figure}
\centering
%\plotancho{./lM.eps}
\includegraphics[width=9.cm,height=8.cm]{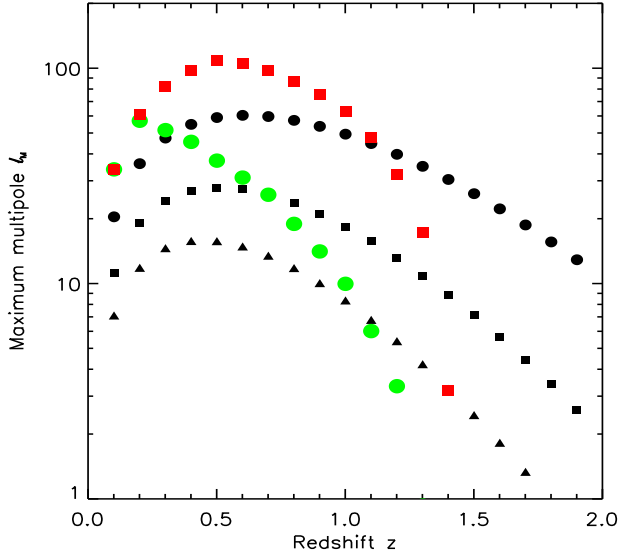}
\caption[fig:lMs]{ Maximum multipole $l_M$ to which a discrete cluster sample on the full sky is sensitive (see equation (\ref{eq:lm})). The ISW is well contained within $l< 30-40$. 
}
\label{fig:lMs}
\end{figure}

%In the context of measuring the blurring term for the CMB intensity, it is obvious that a fundamental error source is the kSZ, i.e., the peculiar motion of the scatterers. In an Einstein de Sitter model, peculiar velocities typically scale with redshift as 
%$v/c \sim 5\times10^{-4}/\sqrt{1+z}$, and drop even more slowly for a universe with cosmological constant. Since the ISW is typically of the order of $\delta T/T_0 \sim 10^{-5}$, we do need a precise subtraction of the kSZ effect, at least in the large scales where the ISW is present. In this sense, this blurring term can be viewed as a correction to the kSZ effect measured in clusters. Depending upon the amplitude of the intrinsic CMB temperature anisotropy at the cluster's position, this correction will range from few percent to 10 --  30 \%, as shown in Figure (\ref{fig:kSZcorr}). 
%We mentioned above that the blurring term keeps memory of what the CMB was like at the cluster's position, and that prompts the possibility of looking at the growth of the ISW by looking at distant clusters. In this section we address the feasibility of such a study.

In the future, several surveys in the optical and in the X-ray range (Pan-STARRS\footnote{Pan-STARRS' URL site: \\
{\tt http://pan-starrs.ifa.hawaii.edu/public/}}, DES \footnote{DES's URL site:\\
  {\tt http://www.darkenergysurvey.org/}}, PAU-BAO\footnote{PAU-BAO's URL
  site: \\ {\tt http://www.ice.csic.es/research/PAU/}} \citep{pau}, Spectrum-X/eROSITA) will probe the cosmological density field up to redshifts $z\sim 0.9 - 1.1$ with unprecedented sensitivity. Gravity relates the matter density distribution with the peculiar motion it gives rise to, so a good estimation for the kSZ in clusters and groups should be obtainable from the density surveys themselves, (this is indeed the goal for cosmological reconstruction algorithms like ARGO, \citet{argo1}). The typical correlation length of the peculiar velocity field is $\sim 40\;$ h$^{-1}$ Mpc (comoving), which at $z\sim 1$ subtends around a couple of degrees. This means that, in the large angular scales where the ISW is present, we should expect to have a fairly high number of uncorrelated estimates of the cluster peculiar velocity. Let us assume that we have, for a given redshift shell, a sample of $N_{cl}$ clusters on the sky. By looking only at the angular position of these clusters, we want to find out  to what range of multipoles we are sensitive. It is clear that, in order to sample a given multipole $l$, our {\it sphere tracers} must lie at distances smaller than $\theta \sim \pi / l$.  For a set of uniformly distributed clusters on the sphere, we can assign an area of $4\pi /N_{cl}$ to each cluster, and hence an average inter-cluster separation of $\theta_{cc} \simeq 2\theta_c = 4/\sqrt{N_{cl}}$  (with $\theta_c = 2/\sqrt{N_{cl}}$ the radius assigned to each cluster area). Therefore, the maximum multipole to which our cluster sample is sensitive is given by 
\begin{equation}
l_M \simeq \frac{\pi}{4} \sqrt{N_{cl}}.
\label{eq:lm}
\end{equation}
Let us know assume that, as justified above, the noise is uncorrelated from pixel to pixel (since the typical separation between pixels corresponds to an actual distance that is larger than the typical correlation length of the kSZ). In this case, we may model the signal in our {\it ith}-pixel as 
\begin{equation}
s(\vnh_i) = s_{CMB}(\vnh_i) + s_{kSZ} (\vnh_i).
\label{eq:sm1}
\end{equation}
We are focusing on the {\em intrinsic} contaminants whose subtraction cannot be improved by additional observations at different frequencies and/or better sensitivities (such as the point source emission or the instrumental noise). Further, the symbol $s_{kSZ}(\vnh_i)$ in the equation above refers to the kSZ {\em residual} that remains after estimating the kSZ (and the peculiar velocity field) from the knowledge of the density distribution in that region of the Universe. We shall see that the requirements on the amplitude of these kSZ residuals are not too stringent, but must be inferred by other observations different to those in the CMB range. We are interested in the signal arising under the cluster's area, and therefore the application of a high-pass filter may be useful to minimize the contamination of the large scale signal.
This filtering should leave some residuals, which are smaller for smaller clusters/groups, and lie typically at the level of 10 -- 30 \% the kSZ amplitude (see Figure (6) in \citet{kszchm}). 

There will be other residuals due to the presence of radio/IR point sources, but, in any case, these residuals share the same statistical properties than the kSZ residuals, and they will be regarded as the same: our goal is to set for them upper limits that enable tracking the growth of the ISW at late epochs. That is, we propose comparing the low $l$ multipoles of the CMB at the high redshift cluster positions with the low $l$ CMB multipoles measured from the whole sky. The difference must be due to the signal that arose between the clusters and the observer, i.e., the ISW.

If clusters are homogeneously distributed over the sky, then it can be easily shown that the error in the estimation of a multipole $a_{l,m}$ in the set of pixels/clusters equals
\begin{equation}
\Delta^2 [a_{l,m}]  \simeq \Omega_{cl} \sigma^2_{kSZ} \simeq \frac{4\pi}{N_{cl}} \sigma^2_{kSZ},
\label{eq:errksz1}
\end{equation}
with $\sigma^2_{kSZ}$ the dispersion of the kSZ residual $s_{kSZ}$ and $\Omega_{cl}$ the solid angle assigned to each cluster. It is possible to make use of the $2l+1$ modes in order to provide an estimate of the power at multipole $l$. The error of such estimate is given by
\begin{equation}
\Delta [C_l]Ê\simeq \frac{\Delta^2 [a_{l,m}]}{2l+1} = \frac{4\pi}{N_{cl}}\frac{\sigma^2_{kSZ}}{2l+1},
\label{eq:errksz2}
\end{equation}
and hence the S/N for the ISW detection at a given multipole reads
\begin{equation}
\left(\frac{S}{N}\right)^2_l = \frac{C_l^{ISW}\; N_{cl}\; (2l+1)}{4\pi\;\sigma^2_{kSZ}}.
\label{eq:errksz3}
\end{equation}
The total S/N is obtained after adding this contribution from $l=2$ up to $l=l_M$ given in equation (\ref{eq:lm}):
\begin{equation}
\left(\frac{S}{N}\right)^2 = \sum_{l=2}^{l=l_M}\left(\frac{S}{N}\right)^2_l.
\label{eq:errksz3}
\end{equation}

%$l$, there are $2l+1$ different modes defining different directions on the sky, and for each direction we need having two pixels/clusters. The number of clusters probing each multipole is approximately $N_{mode,l} = N_{cl} / (2(2l+1))$. Let us assume now that the kSZ introduces some noise in each cluster, of rms $\sigma_v$. If we attempt to measure a given ISW mode of multipole $l$ in the pixel set, the signal to noise will be given by
%\begin{equation}
%\left( \frac{s_l}{n_l}\right)_{mode}^2 = \frac{C_l^{ISW}}{\sigma_v^2 / N_{mode,l}} =
%  \frac{C_l^{ISW} \;N_{cl}}{\sigma_v^2/c^2 \; 2(2l+1)}.
%\label{eq:s2n_mode}
%\end{equation}
%After counting all $(2l+1)$ modes for multipole $l$, we obtain
%\begin{equation}
%\left( \frac{s}{n}\right)_l^2 = \left( \frac{s_l}{n_l}\right)_{mode}^2 (2l+1) = 
% \frac{C_l^{ISW} \;N_{cl}}{2\;\sigma_v^2/c^2},
%\label{eq:s2n_l}
%\end{equation}
%and the total S/N will be given after summing equation (\ref{eq:s2n_l}) from $l=2$ up to the maximum multipole $l=l_M$ given in equation (\ref{eq:lm}),
%\begin{equation}
%\left( \frac{S}{N}\right)^2 = \sum_{l=2}^{l=l_M} \left( \frac{s}{n}\right)_l^2.
%\label{eq:s2nt}
%\end{equation}
 In Figure (\ref{fig:lMs}) we display the maximum $l$ accessible by a cluster population driven from the Sheth-Tormen \citep{ST} mass function in a WMAPV universe with $\sigma_8 = 0.817$. The width of the redshift shell equals $\Delta z = 0.1$. Black circles account for all clusters above $10^{14}$ h$^{-1}$ M$_{\odot}$, whereas black squares and black triangles correspond for $M_{min} = 2, 3 \times 10^{14}$ h$^{-1}$ M$_{\odot}$. We also display the maximum multipoles to be reached with cluster surveys obtained under the nominal sensitivity of the experiment eROSITA:  big green circles correspond to a survey with a flux limit equal to $1.6\times 10^{13}$ erg s$^{-1}$ cm$^{-2}$  in the X-ray band [0.5, 5] KeV ({\it all sky survey}), whereas the big red squares correspond to a flux limit of $3.3\times 10^{14}$ erg s$^{-1}$ cm$^{-2}$ in the same range ({\it wide survey}, see {\tt http://www.mpe.mpg.de/erosita/MDD-6.pdf} for details). We remark that the ISW contain most of its power in the low multipole range, $l<20-30$.\\

\begin{figure}
\centering
%\plotancho{./errorksz.eps}
%\includegraphics[width=6.cm,height=8.cm]{./errorksz.eps}
\plotone{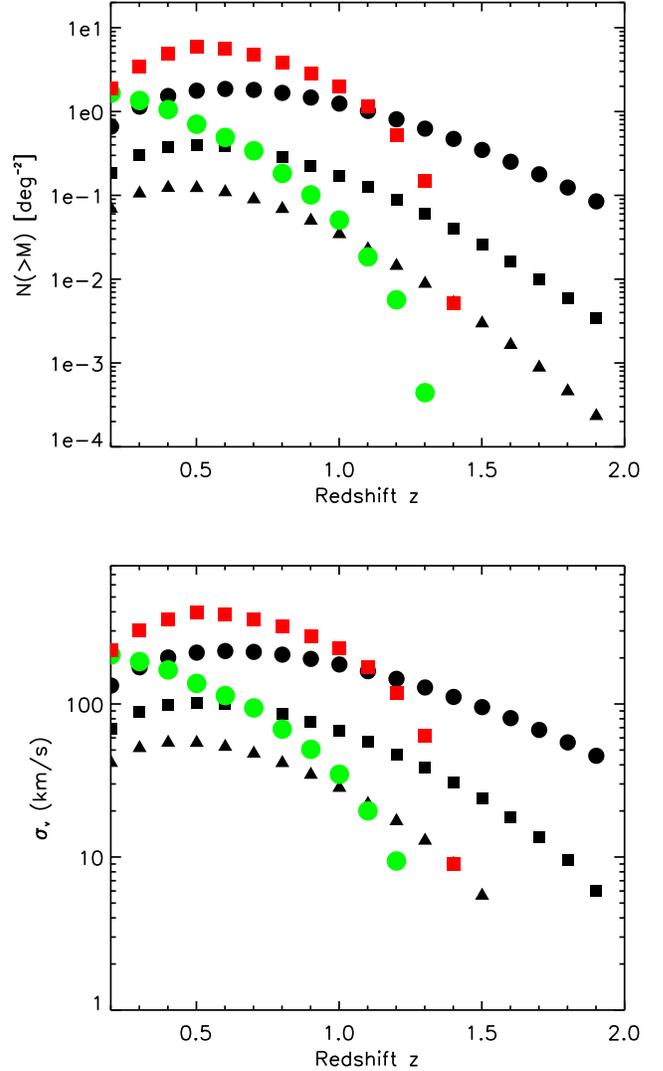}
\caption[fig:errksz]{ {\it Top:} Angular number density of different cluster populations in different redshift shells of width $\Delta z = 0.1$. {\it Bottom:} Required errors in the kSZ recovery in order to obtain residuals 3 times below the ISW level, ($S/N = 3$ in equation (\ref{eq:errksz3}) ).
}
\label{fig:errksz}
\end{figure}

In the top panel of Figure (\ref{fig:errksz}) we display the actual angular density of those cluster populations at different redshifts, again for $\Delta z = 0.1$. In the bottom panel, we display the required error in the peculiar velocity estimates ($\sigma_v = c\;\sigma_{kSZ} $) in order to obtain a residual that is 3 times below the ISW level (i.e., S/N = 3). Symbol coding is the same as in the previous plot. For the cluster sample of $M_{min} = 10^{14}$ h$^{-1}$ M$_{\odot}$ we require errors in the {\em radial }peculiar velocity of the order of 100 -- 200 km s$^{-1}$ in order to see the growth of the ISW at low redsfhits . 
These error requirements  become more stringent when more massive cluster populations are used, and when higher redshifts are to be probed. They are at the level of the actual linear prediction for the kSZ in clusters ($\sim 220$ km s$^{-1}$), so {\it a priori} are not too stringent. However, we must remark that these are the upper limits for the {\em total} errors, and should account for not only kSZ residuals, but also all other types of possible contaminants and systematics. Note as well that the error on the kSZ recovery, $\sigma_v$, is inversely proportional to the required S/N.\\

\section{Constraints on the cosmological dipole}

Let us assume now that we include the dipole in the analyses to be performed at the positions of the cluster distribution. The same requirements that allow tracing the ISW growth should permit, {\it a priori}, to impose constraints on the {\em intrinsic cosmological} dipole of the same order, i.e., at the level of few tens of $\mu$K.  Indeed, if we use equation (\ref{eq:errksz3}) to impose $(S/N)_{l=1} = 1$ with $\sigma_{kSZ} \sim 100$ km s$^{-1}$ and $C_{l=1}^{ISW} \sim 400 (\mu $K$)^2$, we obtain $N_{cl} \sim 4\times 10^3$. This is a relatively modest number of clusters: by including all groups and clusters present in the different redshift shells, the constraints on the cosmological dipole would improve even further and could eventually yield a detection. Already with 4,000 clusters, the constraint of $\sim 20$ $\mu$K is between one and two orders of magnitude below the upper limit of the cosmological dipole that can be inferred from our modelling of the motion of the Local Group. \\

In this context, a natural question that arises is how the subtraction of the measured dipole (which includes both the intrinsic cosmological dipole and the one induced by our peculiar motion) affects our estimates of the intrinsic dipole in the positions of galaxy clusters. Let us assume that the total dipole measured on the {\em whole} sphere is
\begin{equation}
a_{1,m} = a_{1,m}^{intr} + a_{1,m}^{vobs},
\label{eq:dip1}
\end{equation}
built upon components of both the intrinsic cosmological and the observer's velocity dipoles. If, for the time being, we neglect all kSZ residuals in our subset of pixels containing clusters, ideally the total dipole measured at cluster positions would be
\begin{equation}
a^{cl}_{1,m} = a_{1,m}^{intr} (1-\langle \tau_T\rangle) + a_{1,m}^{vobs} ,
\label{eq:dip2}
\end{equation}
but in practice it will be
\begin{equation}
a^{cl,r}_{1,m} \simeq a_{1,m}^{intr} (1-\langle \tau_T\rangle) + a_{1,m}^{vobs} + \sum_{l',m'}\epsilon_{l',m'} a_{l',m'},
\label{eq:dip2b}
\end{equation}
where the last component accounts for the errors introduced by our discrete cluster (pixel) distribution,
and $\langle \tau_T\rangle$ is the average cluster optical depth. Since computing the dipole in our pixel subset is a linear operation on the map, the errors in the dipole estimation will also be linear in the different multipoles conforming the map in the whole sphere. This {\it leakage} of power from different multipoles on the estimated dipole moment $(1,m)$ is described by the geometrical vector $\epsilon_{l',m'}$, which is only dependent on the angular distribution of the cluster set and can be computed with high precision.
%As a first approximation, we shall assume that error in the dipole introduced by our cluster distribution is a geometrical factor $\epsilon_{1,m}$ times the total true dipole in the pixel subset (that is, $a_{1,m}^{cl}$. It is clear that there should be no residuals if the total dipole is zero). 
 Note that we are ignoring the relativistic effects associated to the motion of the observer, \citep{jensgert}. If, as usually done in CMB analysis, the total dipole measured in the entire sphere is subtracted, then, in the cluster pixel set, we would be left with
\begin{equation}
\delta a^{cl}_{1,m} \simeq -\langle \tau_T \rangle a_{1,m}^{intr}  + \sum_{l'>1,m'}\epsilon_{l',m'} a_{l',m'}.
\label{eq:dip3}
\end{equation}
I.e, the rms error on the limits for the intrinsic cosmological dipole will be of the order
\begin{equation}
\left( \Delta [a^{intr}_{1,m}] \right)^{1/2} \sim \frac{\sum_{l'>1,m'}\epsilon_{l',m'} a_{l',m'}}{\langle \tau_T \rangle}.
\label{eq:dip4}
\end{equation}
According to this expression, it becomes critical to choose a convenient cluster/pixel set that fulfills $\epsilon_{l'>1,m'} / \langle \tau_T \rangle \ll 1$, so that interesting limits on $a_{1,m}^{intr}$ can be set. 
If the kSZ removal is accurate and/or the large angle power of the kSZ is negligible, then the residuals 
of our ISW estimation will be dominated by those given in Figure 6 of \citet{kszchm}. In that case, for a population of $\sim 500$ galaxy clusters, the average residual dipole should be of the order of $30 / \sqrt{N_{cl}} \times \sqrt{3} \sim 2 $ $\mu$K if we assume an angular size of 20 -- 30 arcmins, but lower ($\sim 0.4  $ $\mu$K) for smaller sized ($\sim$ 3 arcmin) clusters. These figures ignore the impact of instrumental noise. However,  \citet{sasha_bf} were able to achieve a sensitivity level of a few $\mu$K in each component of the dipole computed on a $\sim 600$ cluster set with WMAP data. A residual dipole amplitude of $\sim 2$ $\mu$K on a cluster set of $\langle \tau_T \rangle \sim 5\times 10^{-3}$ would already correspond to a constraint on the primordial dipole at the level of  $\sim 450$ $\mu$K, which is well below the observed dipole. For the same cluster set, a residual amplitude of $\sim 0.4$ $\mu$K (as quoted for $\sim$ 3 arcmin clusters) should yield a constraint at $\sim 77$ $\mu$K, which is close to the expected value of the intrinsic dipole ($\sim 30$ $\mu$K). We conclude that ongoing CMB experiments like Planck, ACT or SPT should be able to set strong constraints on the intrinsic CMB dipole.

%The geometrical error $\epsilon_{1,m}$ must be known to be smaller than unity for any pixel distribution, so {\it a priori} this approach {\em should enable to search for the cosmological dipole. If the kSZ residuals are such that allow for the characterization of the growth of the ISW, then they should allow for a determination of the intrinsic cosmological dipole at the same expense}.

\section{Discussion and conclusions}

Current CMB high angular resolution surveys like ACT \citep{ACT} or SPT \citep{SPT} are scanning the sky at frequencies in which the identification of galaxy clusters and groups should be possible via the tSZ and the kSZ effects. Nominally, these experiments should be able to detect via the tSZ all clusters more massive than $2\times 10^{14}$ h$^{-1}$ M$_{\odot}$ at large significance, but many smaller cluster and groups should remain close to the detection threshold. Once clusters have been identified by their tSZ distortion, attempts to detect the kSZ in a subset of those systems can be conducted. The kSZ is our only probe for peculiar velocities in the high redshift universe, and those can themselves be used as probes for Dark Energy and missing baryons \citep{kszchm,withho}. In this context, a precise characterization and correction of all possible contaminants is critical. The impact of the blurring effect on the kSZ amounts to $\sim 20$\% for 10\% of the clusters and groups, and becomes more important for 
those objects with small radial projection in their peculiar motions. Provided that the Thomson optical depth of the cluster is known, then this effect should be accurately predicted from background CMB observations (as those available from, e.g., WMAP or Planck). The measurement of this effect is itself a measurement of the gas content of the object under study.

This effect mirrors the CMB intensity at the epoch of scattering, and this is relevant for secondary anisotropies which arise at late times (like the ISW effect): a detection of the blurring effect in objects placed at different redshift shells would provide the picture of the growth of the ISW at recent cosmological times.
 The measurement we are proposing here is statistical, and therefore it does {\em not} require a high S/N in each cluster (just in the same way as in \citet{Eksz} for the kSZ -- E polarization mode cross correlation). Our arguments here are therefore similar to those given in that work: cluster and group positions can be inferred from observations in a wide range of frequencies (optical, IR, X-ray, millimeter), many of which are using those objects for studying the nature of Dark Energy. The critical issue is the nature and the amplitude of residuals in the kSZ/CMB estimation at the cluster positions. To what extent do errors in the IR/radio point source subtraction and/or in the characterization of the local peculiar velocity field actually endanger the ISW measurements? This should critically depend on those errors being systematic or not. If those residuals can be regarded as {\em independent} from cluster to cluster, then the viability of this project should hinge exclusively on the actual precision to which the point source and kSZ residuals can be removed from the clusters' area. Since the ISW contains most of its power at low multipoles ($l_M < $ 10 -- 20), targets may be chosen at a convenient distance in order to minimize the required sky coverage. A possible strategy would then consist on uniformly distributing $N_{patch}$ patches on the sky (with $N_{patch} \sim (4/\pi\; l_M)^2$), lying a distance $\theta \sim 4/\sqrt{N_{patch}}$ away, and scanning deep through each of those patches until finding a set of sources of high S/N at different redshifts. This would improve the efficiency of the survey (since a minimum amount of area would be scanned), at the expense  however of improving the flux/mass thresholds shown in Figure (\ref{fig:errksz}). Hence one would encounter here a trade off between flux sensitive and sky coverage. Whatever approach is finally chosen, it should also provide strong constraints on the cosmological dipole, which would be {\em independent} from those imposed from the local dipole and the local velocity flows.

In this work we propose using, by first time, the so-called {\it blurring term} in Thomson scattering for cosmological purposes. The small fraction of {\em scattered off} CMB photons which are deviated when crossing a galaxy cluster/group should provide information on how the CMB anisotropy field was {\em at the time of the scattering}. If those objects are far enough, then the CMB at that epoch should lack the ISW component that has arisen recently, and this difference could {\it a priori} be picked up after removing all other signals present in the cluster with enough accuracy. Assuming that tSZ residuals are negligible, we find that a typical error of 100 -- 200 km s$^{-1}$ in the cluster peculiar velocity reconstruction is required for all clusters more massive than $10^{14}$ h$^{-1}$ M$_{\odot}$ in order to trace the growth of the ISW at late times. These errors are comparable with the linear expectations for the peculiar motions of those objects,  which involves that {\it (i)} the blurring correction to the kSZ is general of relevance for the estimation of the latter and {\it (ii)} not very precise corrections for the kSZ are required. Those amplitudes roughly correspond to an error of 900 -- 1800 $\tau_T \; \mu$K per cluster. The same level of errors would provide stringent constraints of the intrinsic cosmological dipole. Current and future large scale structure surveys like eROSITA, Pan-STARRS, DES, PAU-BAO, ACT or SPT should already provide enough group and cluster candidates at the relevant redshift ranges. Therefore, the critical point is the feasibility of an accurate  enough kSZ/tSZ/point source subtraction in future high resolution CMB observations.

%The measurement of the polarization at cluster positions should encounter less foregrounds (although its detection requires much more sensitivity {\it per se}). T

%Order of magnitude of the kSZ versus the absorption term.  Argue about Poisson equation and possibility to remove the kSZ. Required number of clusters. For the E component, a local quadrupole caused by peculiar motion of the cluster times a dipole structure in the cluster's surrounding gas distribution.

%\begin{acknowledgements} 

%I
%\end{acknowledgements}

% Bibtex style:------------
\bibliographystyle{aa}
\bibliography{Lit}
%--------------------------

\end{document}